\documentclass[runningheads]{llncs}

\usepackage[T1]{fontenc}
\usepackage{graphicx}
\usepackage{orcidlink}
\usepackage{tikz}
\usepackage{pifont}
\usepackage{tabularx} 
\usepackage{hyperref}
\usepackage{comment}
\usepackage{tcolorbox}
\usepackage{fontawesome5}
\usepackage{cite}

\usepackage{pgfplots}
\pgfplotsset{compat=1.18}

\makeatletter
\renewcommand{\inst}[1]{\leavevmode\unskip$^{#1}$}
\makeatother

\begin{document}
\title{Adoption of Large Language Models in Scrum Management: Insights from Brazilian Practitioners}


\author{%
Mirko Perkusich\inst{1}\orcidlink{0000-0002-9433-4962} \and
Danyllo Albuquerque\inst{1}\orcidlink{0000-0001-5515-7812} \and
Allysson Allex Araújo\inst{3}\orcidlink{0000-0003-2108-2335} \and
Matheus Paixão\inst{2}\orcidlink{0000-0002-1775-7259} \and
Rohit Gheyi\inst{1}\orcidlink{0000-0002-5562-4449} \and
Marcos Kalinowski\inst{4}\orcidlink{0000-0003-1445-3425} \and
Angelo Perkusich\inst{1}\orcidlink{0000-0002-7377-1258}%
}
\authorrunning{Perkusich et al.}
%
\institute{
\inst{1} VIRTUS Research, Development and Innovation Center (UFCG), Brazil \\
\inst{2} State University of Ceará (UECE), Brazil\\
\inst{3} Federal University of Cariri (UFCA), Brazil\\
\inst{4} Pontifical Catholic University of Rio de Janeiro (PUC-Rio), Brazil\\
\email{\{mirko, danyllo.albuquerque, rohit, perkusich\}@virtus.ufcg.edu.br, matheus.paixao@uece.br, allysson.araujo@ufca.edu.br, kalinowski@inf.puc-rio.br}
}
\maketitle              
%

\begin{abstract}
Scrum is widely adopted in software project management due to its adaptability and collaborative nature. The recent emergence of Large Language Models (LLMs) has created new opportunities to support knowledge-intensive Scrum practices. However, existing research has largely focused on technical activities such as coding and testing, with limited evidence on the use of LLMs in management-related Scrum activities. In this study, we investigate the use of LLMs in Scrum management activities through a survey of 70 Brazilian professionals. Among them, 49 actively use Scrum, and 33 reported using LLM-based assistants in their Scrum practices. The results indicate a high level of proficiency and frequent use of LLMs, with 85\% of respondents reporting intermediate or advanced proficiency and 52\% using them daily. LLM use concentrates on exploring Scrum practices, with artifacts and events receiving targeted yet uneven support, whereas broader management tasks appear to be adopted more cautiously. The main benefits include increased productivity (78\%) and reduced manual effort (75\%). However, several critical risks remain, as respondents report ‘almost correct’ outputs (81\%), confidentiality concerns (63\%), and hallucinations during use (59\%). This work provides one of the first empirical characterizations of LLM use in Scrum management, identifying current practices, quantifying benefits and risks, and outlining directions for responsible adoption and integration in Agile environments.


\end{abstract}

\section{Introduction}
\label{sec1-intro}

Agile project management has increasingly shaped contemporary software engineering practices as teams seek ways to handle evolving requirements, coordinate work more effectively, and maintain consistent delivery rhythms \cite{shastri2021spearheading}. Within this context, Scrum has emerged as the most widely adopted framework, structuring teamwork through artifacts, events, and roles that facilitate coordination and continuous delivery~\cite{verwijs2023theory, DigitalAI2025StateOfAgile}. These practices are highly knowledge-intensive, requiring effective communication, documentation, and stakeholder alignment \cite{burga2022examining}. In particular, the Brazilian software industry has seen a substantial increase in Agile maturity, with companies adopting Scrum to manage increasingly complex, fast-paced projects across sectors such as fintech, e-commerce, and public-sector digital services \cite{ABES2025mercado}. This growth intensifies the need for practices that support communication and knowledge work, positioning Brazil as an important context for understanding how LLMs are being incorporated into Scrum management.

At the same time, the rapid evolution of Large Language Models (LLMs)—accessible through platforms such as ChatGPT, Gemini, Claude, and DeepSeek—has introduced new opportunities to support Scrum teams' activities. In fact, LLMs have already demonstrated an impact in technical areas of software engineering, including requirements definition \cite{arora2024advancing}, architecture design \cite{jahic2024state}, coding \cite{fan2025exploring}, debugging \cite{lee2024github}, testing \cite{mathews2024test}, and documentation \cite{hou2024large}. These capabilities fuel interest in understanding how LLMs may extend their influence to management activities. However, agile management activities (such as refining artifacts, facilitating events, and supporting role-specific responsibilities) pose distinct challenges because they involve ambiguity, negotiation, prioritization, and human dynamics \cite{kerzner2025project}. These dimensions, central to Scrum, demand contextual understanding and shared meaning-making \cite{alami2022scrum}. As Agile practices mature, understanding how LLMs can support the management dimension of Scrum becomes increasingly relevant \cite{karnouskos2024relevance}. However, the literature concentrates on assistance for technical tasks, leaving open questions about how LLMs are being applied to artifact refinement, event preparation, or communication among Scrum roles \cite{dhruva2024agile}. In other words, organizations risk adopting these tools based on ``hype'' rather than evidence. This perspective includes challenges such as over-reliance on AI, propagation of biased outputs, confidentiality risks, and potential erosion of shared team accountability.

To address the aforementioned challenges, this study aims to understand how software professionals are incorporating LLMs into Scrum management activities. In particular, we focus on Brazilian professionals because Brazil hosts one of the largest and fastest-growing software markets in Latin America, where Scrum has become the predominant approach for structuring digital transformation initiatives in both private and public organizations \cite{ABES2025mercado, fontana2022countrywide}. To examine these practices at scale and generate an initial characterization anchored in practitioners’ real-world behavior, we adopt a survey as our research method. Surveys enable the identification of broad usage trends, perceived benefits, and concerns across heterogeneous organizational settings, offering the empirical breadth needed to study a phenomenon that is still emerging and scarcely documented in the literature \cite{molleri2016survey}.


Regarding the contributions, this study offers three insights into how LLMs are being incorporated into Scrum management activities. First, we map how Brazilian professionals employ LLM-based assistants to support artifacts, events, and role-specific responsibilities. Second, we identify perceived benefits (such as productivity gains, improved documentation, enhanced communication, and support for decision-making) that illustrate where LLMs already add value to Agile work. Finally, we expose the main barriers influencing adoption, including trust issues, confidentiality risks, unreliable outputs, and organizational resistance. 


The remainder of this paper is organized as follows. Section~\ref{sec2-rel_work} reviews related work. Section~\ref{sec3-design} describes the research design. Section~\ref{sec4-results} presents the results, and Section~\ref{sec5-implications} discusses their implications. Section~\ref{sec6-threats} outlines threats to validity, and Section~\ref{sec7-final} concludes the paper.

\section{Related Work}
\label{sec2-rel_work}

This section briefly reviews prior work across three areas relevant to this study: (i) the use of LLMs in software engineering tasks, including their capabilities and known limitations; (ii) the adoption of Artificial Intelligence (AI) assistants in professional and organizational settings; and (iii) research on Agile management, particularly Scrum. By connecting these strands, we also highlight the current lack of empirical evidence on how LLMs are used in real Scrum management activities, which motivates our investigation.

\textbf{\textit{LLMs in Software Engineering}}. LLMs have become transformative assets in software engineering, supporting technical activities across the entire development lifecycle, including requirements elicitation, code generation, testing, maintenance, documentation, and analytics \cite{Viet2024Large}. Different surveys already reinforce this rapid expansion: \cite{hou2024large} reviewed nearly 400 studies spanning multiple SE phases, while \cite{Zhang2023Survey} mapped research clusters showing strong growth in LLM-assisted coding, testing, and defect prediction. Despite these advances, significant challenges remain. Reliability issues (such as hallucinations and inconsistent outputs) have been widely documented \cite{Gao2025Current}, and the proliferation of benchmarks \cite{Hu2025Assessing} highlights ongoing difficulties in evaluation and standardization. Beyond code-centric tasks, LLMs have shown promise in supporting complex and knowledge-intensive activities, such as software model evolution \cite{Tinnes2023Towards}. However, their role within broader organizational and managerial settings is still emerging.

\textbf{\textit{LLMs in Professional and Organizational Contexts}}. LLMs are increasingly integrated into professional environments to enhance productivity, creativity, and decision-making. Studies demonstrate their ability to support workplace choices by providing rationale and recommendations \cite{Kim2024LeveragingLL}, as well as to assist group work by analyzing meeting transcripts and enhancing coordination \cite{Lubos2025TowardsGD}. They also strengthen information management through more efficient query resolution and document summarization, contributing to higher work productivity \cite{Agarwal2024HarnessingTP}. However, their performance varies across cognitive tasks: while LLMs can mimic human-like reasoning \cite{Tang2024HumanlikeCP}, they still struggle with creativity and context-sensitive ideation. In specialized domains, professionals find them more effective for translation and review than for planning or generating novel concepts \cite{Chakrabarty2024CreativitySI}. Broader surveys indicate that knowledge workers mostly rely on LLMs for code generation, refinement, and text improvement, while expecting deeper enterprise integration in the near future \cite{10.1145/3613905.3650841}. Successful adoption depends on trust, transparency, and skillful prompt engineering \cite{Eigner2024DeterminantsOL}. Although these studies highlight the benefits of LLMs in individual and group productivity, their implications for collaborative, knowledge-intensive methods, such as Agile and Scrum, remain insufficiently understood.

\textbf{\textit{Agile project management}}. Agile marked a shift from hierarchical to collaborative and knowledge-intensive practices \cite{Fernandez2008AgilePM}. Prior studies emphasized that agile teams rely on tacit knowledge exchange and personalization strategies rather than codification, with up to 81\% of knowledge-management activities occurring through social interaction \cite{Vasanthapriyan2019KnowledgeMI, Ouriques2018KnowledgeMS}. Agile practices, such as pair programming, Scrum meetings, and continuous stakeholder involvement, foster transparency, learning, and a shared understanding \cite{Singh2015KnowledgeMT}. Organizational culture is a critical enabler of effective knowledge management within agile teams, shaping collaboration and sustaining productivity \cite{Paterek2016ProjectMD}.

Building on these empirical foundations, emerging research has explored whether LLMs can support or extend these knowledge-intensive dynamics. Early studies experimented with LLMs acting as Scrum Masters in educational settings \cite{couder2024large}, automating project-planning tasks through systems such as AutoScrum \cite{schroder2023autoscrumautomatingprojectplanning}, or coordinating software-process simulations via multi-agent models \cite{Lin2024SOEN101CG}. Other work investigated LLMs’ ability to extract domain models from agile backlogs \cite{Arulmohan2023ExtractingDM}. While these results are promising, they remain largely confined to prototypes and academic scenarios, leaving open the question of how LLMs are actually being adopted in real-world Scrum management activities.

\textbf{\textit{Research Gap}}. LLMs have demonstrated potential across various software engineering tasks, indicating their value in knowledge-intensive activities. Scrum, as a highly collaborative and adaptive framework, represents promising environment for using such tools, including in backlog refinement, event preparation, and support to role-specific responsibilities. Existing studies have begun to explore these possibilities, but most remain confined to prototypes or educational scenarios, offering limited visibility into professional practice. Despite growing interest, there is still a lack of empirical evidence on how LLMs are being adopted within real Scrum teams, what benefits practitioners perceive, and which challenges or risks emerge. To address this gap, this study examines the use of LLMs in Scrum management activities, aiming to map current practices, benefits, and barriers from the perspective of Brazilian practitioners.

\section{Research Design}
\label{sec3-design}

\begin{figure}[!h]
  \centering
  \includegraphics[width=0.8\linewidth]{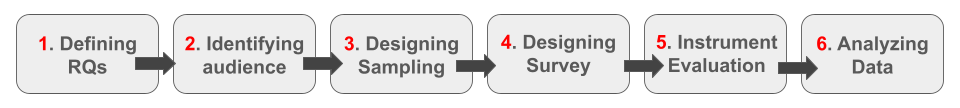}
  \caption{Methodological workflow of the questionnaire survey.}
  \label{fig:method-workflow}
\end{figure}

This study followed a systematic process for conducting a survey in software engineering, drawing on the recommendations of Wagner \textit{et al}. \cite{wagner2020challenges}. Figure~\ref{fig:method-workflow} provides an overview of this process, which was organized into six key steps described below.

\textbf{\textit{1) Defining Research Questions (RQs)}}. We followed a \textit{Goal-Question-Metric} (GQM)-driven survey design~\cite{kalinowski2024teaching} and defined the RQs to capture different dimensions of how LLMs are being used in Scrum management activities. Each RQ (see Table \ref{tab:RQs}) addresses a specific gap identified in the literature and practice.

\begin{table}[!h]
\centering
\caption{Research Questions, Descriptions, and Rationales.}
\label{tab:RQs}
\small
\scalebox{0.75}{
\begin{tabular}{p{1.0cm} p{6.3cm} p{10.3cm}}
\hline
\textbf{RQ} & \textbf{Goal} & \textbf{Motivation} \\
\hline

\textbf{RQ1} &
What is the current level of knowledge and usage of LLMs among Scrum professionals? &
Understanding baseline familiarity and adoption is essential because the diffusion of innovations depends not only on tool availability but also on practitioners' awareness, proficiency, and willingness to integrate them into management-oriented activities. \\

\textbf{RQ2} &
In which Scrum practices (artifacts, events, and roles) are LLMs being adopted, and how helpful are they perceived to be? &
Scrum is structured around artifacts, events, and roles. Identifying where LLMs already contribute helps reveal how these tools support real management work, rather than only technical tasks. \\

\textbf{RQ3} &
What benefits do Scrum practitioners perceive from using LLMs? &
Industry narratives emphasize the potential of LLMs, but empirical evidence is limited. Mapping perceived benefits clarifies whether LLMs provide tangible value in practice, such as increased productivity, improved documentation quality, or enhanced decision-making support. \\

\textbf{RQ4} &
What risks and challenges are associated with LLM use in Scrum contexts? &
LLMs can produce biased or incorrect outputs, creating risks such as over-reliance, privacy concerns, and resistance from team members. Understanding these barriers is crucial to avoiding naïve adoption and to guiding responsible use in Agile environments. \\

\hline
\end{tabular}}
\end{table}

These RQs provide a comprehensive framework for investigating the role of LLMs in Scrum management activities. By combining descriptive findings on knowledge and adoption (RQ1), practical applications in Scrum practices (RQ2), perceived benefits (RQ3), and experienced risks and challenges (RQ4), the study aims to build a balanced understanding of both the opportunities and limitations of LLMs.

\textbf{\textit{2) Identifying Target Audience and Sampling Frame}}. The target audience of this study comprised professionals with practical experience in Scrum-based projects. To ensure respondent eligibility, a screening question required participants to confirm prior or current participation in Scrum projects; those without such experience were automatically redirected and excluded, in accordance with recommended practices for survey validity \cite{kasunic2005designing}. The sampling frame was intentionally broad, aiming to capture the diversity of the Scrum community across industries, organizational sizes, and team roles, including Product Owners, Scrum Masters, developers, and testers. Although this heterogeneity enhances ecological validity, it does not yield a statistically representative sample, an inherent limitation of non-probabilistic sampling in surveys \cite{molleri2020empirically}.

\textbf{\textit{3) Designing Sampling Plan}}. To operationalize the sampling frame, we employed a non-probabilistic strategy combining convenience and snowball sampling, a common approach in software engineering surveys where practitioner registries are unavailable \cite{dillman2014internet}. Recruitment occurred through multiple channels (e.g., LinkedIn Agile communities, professional mailing lists, and direct invitations to industry and academic contacts), with respondents encouraged to share the survey to broaden reach. Although the initial goal was to obtain at least 70 valid responses for descriptive and comparative analyses \cite{graziotin2021psychometrics}, participation remained voluntary. Following Dillman’s tailored design principles, the invitation emphasized clarity of purpose, approximate completion time, and full anonymity, aiming to reduce response bias and increase engagement \cite{dillman2014internet}.

\textbf{\textit{4) Designing Survey Instrument}}. The instrument was structured into nine sections, each aligned with the study’s RQs and grounded in established guidelines for survey-based research \cite{molleri2020empirically}\cite{kasunic2005designing}. The instrument consisted primarily of closed-ended items (multiple-choice questions, Likert scales, and frequency measures) with a small number of open-ended questions to capture qualitative insights and provide illustrative examples. Operationalized in Google Forms, the survey was voluntary, anonymous, and designed to strike a balance between breadth of coverage and ease of completion. Table~\ref{tab:instrument} summarizes the structure of the instrument and its connection to the RQs. The complete questionnaire used in this study is provided in the supplementary material (see \nameref{sec-artifact}).

\begin{table}[!h]
\centering
\caption{Survey sections, descriptions, and relation to RQs.}
\label{tab:instrument}
\small
\scalebox{0.75}{
\begin{tabular}{p{4.8cm} p{10.5cm} p{1.5cm}}
\hline
\textbf{Section} & \textbf{Description} & \textbf{RQs} \\
\hline
1–3: Profile definition & Collects demographic and background data about participants (country, age, education), organizational context (industry, size, team), and project characteristics (role, domain, Scrum adoption). & - \\

4: Knowledge and use of LLMs & Assesses familiarity with LLMs, adoption mode (formal/informal), frequency of use, time spent, and models/versions employed. & RQ1 \\

5: Use in Scrum & Investigates LLM use across Scrum artifacts, events, and roles, including perceived helpfulness and potential role replacement. & RQ2 \\

6: Perceived benefits & Identifies benefits already experienced with LLMs (e.g., productivity, quality, decision-making, and communication) and collects positive examples. & RQ3 \\

7: Difficulties and challenges & Explores encountered difficulties (e.g., bias, privacy, resistance, over-dependence), agreement with risk statements, and negative examples. & RQ4 \\

8: General comments & provides an open space for additional insights, suggestions, or reflections about LLMs in Scrum. & RQ1–RQ4 \\

9: Closure & Final thanks, option for follow-up contact, and confirmation of survey completion. & – \\
\hline
\end{tabular}}
\end{table}

\textbf{\textit{5) Survey Instrument Evaluation}}. The questionnaire underwent a multi-step evaluation process to ensure clarity, relevance, and alignment with the study objectives, following survey research guidelines in software engineering \cite{graziotin2021psychometrics,molleri2020empirically}. First, three researchers with expertise in empirical SE, Agile methods, and AI reviewed the initial version of the instrument, refining wording, removing redundancies, harmonizing Scrum terminology, clarifying key concepts (e.g., LLM adoption modes), and reordering sections to improve flow and content validity. Next, the instrument was piloted with five Scrum practitioners (PO, SM, and developers) from different organizations, who completed the form and provided structured feedback on clarity, terminology, layout, and timing; this led to minor adjustments such as simplified instructions, reduced scale redundancy, more consistent terminology, and improved sequencing, resulting in an average completion time of about 10 minutes. To ensure data quality, the questionnaire included simple attention checks and automatic routing that redirected respondents without Scrum or LLM experience to a closure section, minimizing careless responses and enforcing eligibility in line with \cite{dillman2014internet}. Ethical procedures included an informed consent statement describing the study’s aims, risks, voluntariness, and anonymity; no identifiable data were collected, and all data were stored securely and accessed only by the research team. Overall, this process enhanced the clarity, reliability, and validity of the instrument, supporting high-quality data collection and future replication.

\textbf{\textit{6) Analyzing Survey Data}}. The questionnaire was deployed through Google Forms and remained open throughout October and November 2025. Participation was voluntary and anonymous, with informed consent and automated routing to exclude respondents without Scrum experience or prior use of AI chat assistants. The response flow was periodically monitored to ensure accessibility and prevent technical issues, and no personally identifiable information was collected. Because most items were mandatory, the resulting dataset contained no missing values. After collection, all responses were exported to \texttt{XLSX} format for further processing.

Data analysis combined quantitative and qualitative techniques. Closed-ended items were examined through descriptive statistics such as frequencies, percentages, and means, while open-ended responses were coded using a mixed deductive–inductive coding approach. To improve the reliability of qualitative interpretations, two researchers independently coded a subset of responses and reconciled discrepancies through discussion. Analyses were explicitly mapped to the research questions: RQ1 addressed participant profiles and familiarity with LLMs; RQ2 examined usage across Scrum artifacts, events, and roles; RQ3 focused on perceived benefits; and RQ4 investigated risks and challenges. All procedures followed established guidelines for survey research in software engineering \cite{graziotin2021psychometrics, dillman2014internet,molleri2020empirically}. The dataset, survey instrument, and analysis scripts are available as supplementary material (see the \nameref{sec-artifact} section).

\section{Results and Discussion}
\label{sec4-results}

This section presents and discusses the survey's results. The first part characterizes the respondents, their organizations, and project contexts. The subsequent subsections integrate findings and interpretative discussion for RQ1–RQ4. Additional breakdowns, supporting tables, and the complete dataset are available in the supplementary material (see the \nameref{sec-artifact} section).

\subsection*{Participant, Organizational, and Project Profiles}


A total of 70 Brazilian professionals completed the demographic section. Of these, 49 had worked on a Scrum project in the past six months and reported using AI chat assistants in Scrum-related activities; therefore, they completed the full instrument. The analyses presented in the following focus on this sample.

The demographic profile indicates a predominantly young respondent base: 83\% are between 20 and 39 years old, 14\% fall in the 40–49 range, and only 2\% are older than 50. Formal education levels are high, with 63\% holding undergraduate degrees or MBAs and 20\% reporting a master’s or doctoral degree. Regarding Scrum knowledge, 69\% classify themselves as ``Beginner'' (knowledge of the main practical aspects) or ``Qualified'' (solid conceptual understanding and practical experience), reflecting a population with well-established familiarity with Agile practices. This educational and professional profile aligns with the strong presence of technical roles in the sample.

Respondents work across diverse industries, although technology-oriented sectors dominate. Technology accounts for the largest share (25\%), followed by education (18\%) and research, development, and innovation (16\%). Smaller proportions come from finance, retail, industry, public administration, and consulting. Organizational size also varies substantially: 33\% come from companies with more than 1,000 employees, 27\% from medium-sized companies (251–1,000 employees), and the remaining 40\% from small organizations with up to 250 employees. This distribution indicates that Scrum practices and AI-supported workflows are present in both consolidated enterprises and smaller, fast-moving companies.

Regarding professional roles, developers constitute the largest segment of the sample (47\%), followed by Scrum Masters (12\%), QA/testers (8\%), and Product Owners (6\%). In terms of application domains, Web and cross-platform development predominate (66\%), with additional representation from education (21\%), finance (18\%), transport and distribution (16\%), and telecommunications (10\%). Most initiatives involve small teams of 1–10 members (65\%), and over half of the projects (53\%) have been running for one to three years. 



\subsection*{Knowledge and Usage of LLMs Among Scrum Professionals (RQ1)}

RQ1 examined how familiar Brazilian Scrum professionals are with AI chat assistants and how frequently they use them. The results reveal a user base that is already relatively mature in its understanding of LLMs. Among the respondents who use AI assistants, 61\% describe themselves as ``Qualified'' and 21\% as ``Proficient''. Only 15\% classify themselves as ``beginners''. Usage frequency follows a similarly intensive pattern. Daily use is reported by 52\% of respondents, weekly use by another 18\%, and monthly use by 21\%. Only 9\% indicate sporadic use, and no respondent reports having abandoned the technology after initial experimentation. Time allocation also reflects sustained integration: among valid answers, 47\% use assistants for less than one hour per day, while another 50\% use them between one and two hours daily. 

The ecosystem of tools and models used by respondents reveals a diversified but strongly centralized pattern. ChatGPT is universally adopted (100\%), followed by Gemini (82\%) and Copilot Chat (67\%). Adoption of Claude and DeepSeek reaches only 27\%, while niche tools such as Perplexity, Manus, Qwen, Phind, and Grok appear in less than 12\% of responses, reflecting exploratory rather than systematic use. At the model level, references are similarly concentrated: GPT-4o (27\%), GPT-5/5o/5.1 reported by 21\% when aggregated, Gemini 2.5 Pro (21\%), and Gemini 2.5 Flash (15\%) were the most cited. Mentions of Claude 4.1, Sonnet, Llama, and Qwen were isolated, showing that advanced model awareness is present but uneven across practitioners. 

Interaction patterns emphasize ``text'' as the default modality (100\%), but document-based workflows (76\%) and image/screenshot analysis (61\%) are already common, highlighting multimodal adoption. More than half the sample (55\%) uses LLMs in technical tasks (debugging, coding, testing), reinforcing the dual role of assistants in both managerial and technical activities within Scrum practitioners. Organizational governance is mixed: 36\% report no policy, 36\% have formal policies, and 12\% rely on informal guidelines, suggesting fragmentation and a lack of institutional standardization. Despite this, 51\% use LLMs informally, while 45\% follow a prescribed process, reinforcing that adoption is practitioner-driven rather than organizationally mandated.

\begin{tcolorbox}[colback=gray!5, colframe=black!50, title=\faLightbulb\quad \textbf{Summarized Answer - Knowledge and Usage of LLMs (RQ1)}, rounded corners,  arc=3pt, boxsep=2pt,
  boxrule=0.4pt]
\scriptsize
The results indicate that LLM use among Brazilian Scrum practitioners is routine rather than exploratory. Practitioners rely on LLM assistants across multiple modalities, integrating them into both managerial and technical tasks through short, iterative interactions. Adoption is concentrated in major ecosystems (ChatGPT, Gemini, and Copilot), with some experimentation involving smaller models. However, organizational governance has not kept pace: both the presence of formal policies and the complete absence of policies account for 36\% of responses, indicating practitioner-driven adoption. Overall, the findings suggest that LLMs are embedded as everyday cognitive aids for reasoning and problem-solving, while core responsibility and decision-making remain human-centered.
\end{tcolorbox}


\subsection*{Adoption Across Scrum Practices (RQ2)}

\begin{figure}[!ht]
\centering
\includegraphics[width=0.95\linewidth]{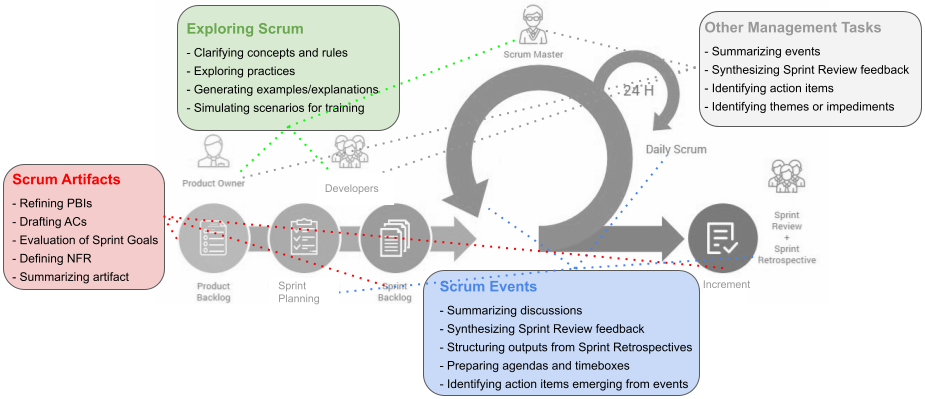}
\caption{LLM uses across Scrum activities: learning, artifacts, events, and management tasks.}
\label{fig:scrumllm}
\end{figure}

RQ2 examined where, across Scrum management activities, LLM chat assistants are being used and how helpful practitioners perceive them.
To contextualize these results, Figure \ref{fig:scrumllm} first overlays the classical Scrum workflow with the main areas where LLMs support practitioners, such as learning Scrum, refining artifacts, assisting events, and complementing broader management tasks. This figure provides a structural view of how LLM support aligns with the framework's canonical elements.

\begin{figure}[!ht]
\centering
\includegraphics[width=\linewidth]{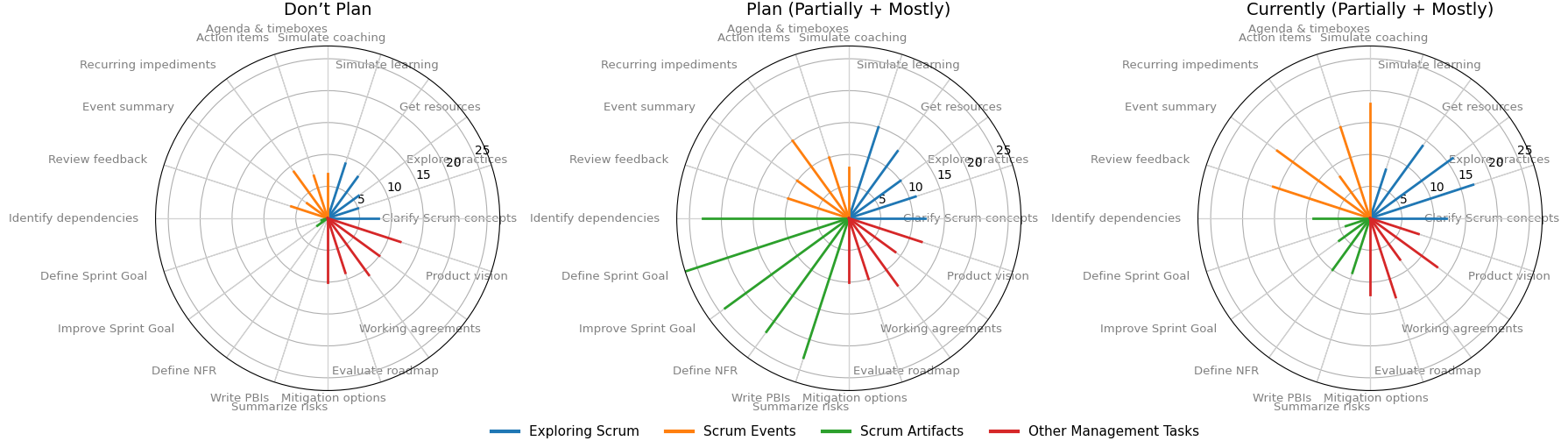}
\caption{Non-planned, planned, and current use of LLMs across categories.}
\label{fig:radar}
\end{figure}

Across all categories, respondents indicate that LLM assistants are used primarily as supportive cognitive tools rather than as substitutes for human decision-making. Their use clusters around four major areas. As shown in Figure \ref{fig:radar}, these domains exhibit distinct adoption profiles, both in terms of current use and future intention to use. ``Exploring Scrum'' dominate the outer rings, reflecting widespread use. ``Artifact and event'' items cluster in the mid-ring, showing substantial but selective adoption. Finally, ``other Management tasks'' appear closer to the inner rings, confirming cautious use.

\textbf{Exploring Scrum} exhibits the most consistent adoption. For all five exploration activities, the respondents report either ``current use'' or ``plans to use'' LLMs. Exploring practices (82\% use or plan), practical resources (78\% use or plan), and clarifying Scrum concepts (72\% use or plan) are the most frequently adopted. In all cases, current partial use is the dominant pattern (22–35\%), confirming that LLMs have become a primary cognitive aid for understanding, explaining, and preparing Scrum work. 

LLM support for \textbf{Scrum Artifacts} also exhibits substantial adoption. Core refinement tasks (writing Product Backlog Items (PBIs), defining Non-Functional Requirements (NFRs), improving Sprint Goals) show 72–87\% reported ``use or planned use''. For instance, 77\% intend to or currently use LLMs to improve Sprint Goals (19 ``plan mostly'', 4 ``currently partially'', and 2 ``currently mostly''), and 76\% do so for defining NFRs. These activities align closely with LLMs’ strengths in text transformation, summarization, and structured reasoning, which explains their high levels of acceptance.

In \textbf{Scrum Events}, adoption is led by summarization and synthesis activities. Summarizing what happened during events reaches 68\% combined use/plan (with 36\% already using), while synthesizing Sprint Review feedback reaches 64\% (with 42\% already using). More interpretive tasks show lower adoption: designing agendas/timeboxes (54\%) or identifying impediments (48\%) involve judgment and contextual nuance, explaining the more cautious uptake. Nonetheless, even these tasks show non-trivial planned adoption, indicating growing trust in LLM-assisted facilitation. 

Finally, Adoption for \textbf{Other Agile Management Tasks} is more heterogeneous. Moderate levels of ``use or planned use'' (30–45\%) appear in summarizing key risks (47\%) and defining mitigation options (43\%), reflecting practitioners’ interest in LLM-based analytical support. However, alignment-heavy or culturally embedded tasks show clear resistance: improving working agreements (33\% use/plan, 33\% don’t plan) and describing product vision (29\% use/plan, 38\% don’t plan). These results underscore that practitioners intentionally retain human ownership over value-driven and consensus-sensitive activities.

Overall, Figure~\ref{fig:radar} indicates a consistent pattern in which practitioners rely on LLMs for text-heavy, analytical, and exploratory tasks, while responsibilities requiring judgment, collaboration, or strategic reasoning continue to be handled primarily by humans. This finding points to a pragmatic form of hybrid human–AI work within Scrum practitioners.

\begin{tcolorbox}[colback=gray!5, colframe=black!50, title=\faLightbulb\quad \textbf{Summarized Answer - Adoption Across Scrum Practices (RQ2)},  rounded corners,  arc=3pt, boxsep=2pt,
  boxrule=0.4pt]
\scriptsize
The results indicate that Brazilian Scrum practitioners widely use LLM assistants in management activities, primarily as supportive aids rather than substitutes for human judgment. Conceptual exploration is the most common use, while Scrum events mainly involve summarization and synthesis, with planning tasks used more cautiously. Artifact-related work shows the highest acceptance, reflecting strong alignment between LLM capabilities and text-intensive refinement tasks. In broader management activities, adoption is uneven, with summarization and risk-related uses gaining traction, while vision-setting and team alignment face greater resistance.
\end{tcolorbox}


\subsection*{Perceived Benefits (RQ3)}

\begin{figure}[!h]
    \centering
    \includegraphics[width=0.85\linewidth]{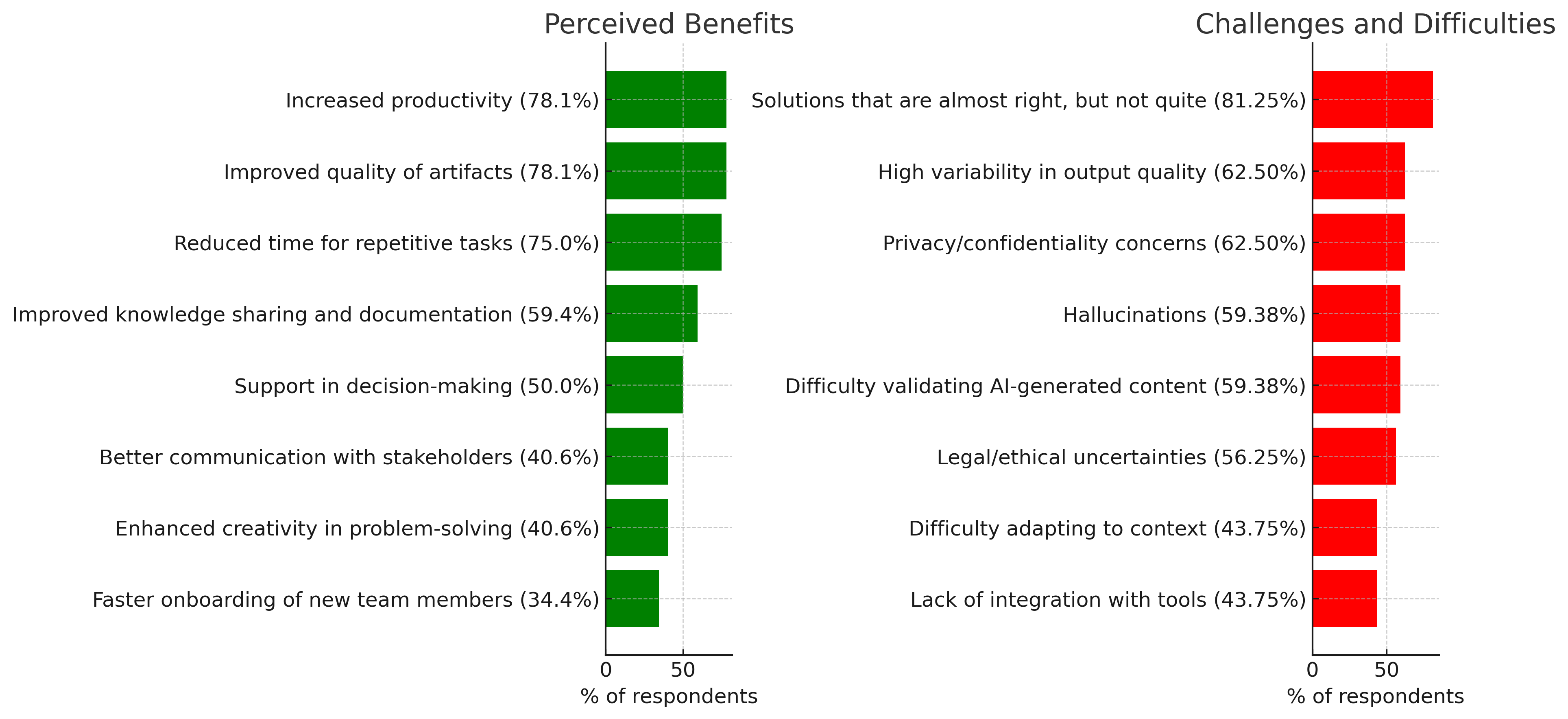}
    \caption{Perceived benefits and challenges associated with the use of AI-based assistants.}
    \label{fig:benefits-challenges}
\end{figure}

Figure \ref{fig:benefits-challenges} summarizes the benefits most frequently reported by practitioners. The distribution reveals that AI chat assistants are perceived as highly effective in enhancing individual efficiency, improving artifact quality, and reducing repetitive cognitive load. Productivity and quality improvements (each 78\%) and reduced effort in repetitive tasks (75\%) form a cluster of strong, text-intensive benefits, suggesting that LLMs have become reliable companions for refinement, rewriting, summarizing, and structuring Scrum-related information.

Beyond speed and quality, respondents also highlighted benefits tied to knowledge work. Improved documentation and knowledge sharing (59\%) reflect that AI assistance is increasingly embedded in organizational memory practices. Half of the respondents also report support for decision-making, indicating that LLMs are used not only as writing tools but also as sensemaking partners in everyday reasoning tasks. Collaboration-oriented gains (improved stakeholder communication and enhanced creativity) appear less frequently ($\approx$40\%) yet still point to emerging uses where LLMs help practitioners articulate ideas, explore alternative solution paths, and communicate more effectively.

Responses to benefit-oriented statements further reinforce these trends. Most participants agree that AI assistants help them work more efficiently (89\%) and reduce cognitive effort in routine activities (85\%). Perceived team-level improvements are also strong (83\%). However, benefits tied to collaboration (64\%) and transparency or traceability (57\%) are more modest, suggesting that organizations are still navigating how to consistently integrate AI-generated content into workflows, governance mechanisms, or shared documentation.

We also found that perceived usefulness varies across Scrum roles. Developers report overwhelmingly positive perceptions (84\% finding AI useful or very useful), reflecting strong alignment with task-level reasoning, refinement, and technical exploration. Scrum Masters show moderate but steady benefits (56\%), consistent with facilitation, coaching, and impediment-resolution tasks that can benefit from AI but remain deeply human-centered. Product Owners display the most heterogeneous perceptions: while some find LLMs useful for prioritization, vision communication, or backlog reasoning, a notable portion reports low applicability, reflecting the strategic nature of PO responsibilities.


\begin{tcolorbox}[colback=gray!5, colframe=black!50, title=\faLightbulb\quad \textbf{Summarized Answer - Perceived Benefits (RQ3)},  rounded corners,  arc=3pt, boxsep=2pt,
  boxrule=0.4pt]
\scriptsize
LLM assistants are widely perceived as valuable aids in Scrum management, particularly for efficiency, quality, and cognitive relief. Practitioners report substantial gains in productivity, artifact refinement, and reduction of repetitive effort, with strong perceptions of improved individual and team performance. Benefits involving collaboration and transparency appear more moderate, reflecting partial and uneven integration into collective workflows. Perceived usefulness differs across roles: Developers report the strongest benefits, Scrum Masters show steady but moderate gains, and Product Owners express mixed views. Overall, LLMs reinforce text-intensive and reasoning-focused tasks, while their impact on coordination or strategic responsibilities remains more situational.
\end{tcolorbox}

\subsection*{Risks and Challenges (RQ4)}

Figure~\ref{fig:benefits-challenges} also highlights the challenges and frustrations practitioners face when using AI chat assistants for Scrum work. The most frequently cited difficulty is receiving outputs that are ``almost correct but not quite'' (81\%), which aligns with known limitations of LLMs in precision-required contexts. High variability in output quality (63\%) further reinforces the perception that AI responses require continuous monitoring and cannot yet be fully trusted for autonomous use.

In summary, we found that trust-, safety-, and reliability-related issues are central to the reported challenges. Privacy and confidentiality concerns (63\%), hallucinations (59\%), and difficulty validating AI-generated content (59\%) illustrate that practitioners must apply active oversight, especially in contexts involving sensitive information or formal decision-making. Legal and ethical uncertainties (56\%) additionally underscore unresolved organizational questions surrounding responsible use, compliance, and governance. Contextual adaptation issues---difficulty tailoring outputs to team or project realities (44\%)---and lack of tool integration (44\%) reveal frictions in practical adoption. These results show that although LLMs provide substantial value, their usefulness is constrained by organizational readiness, ecosystem integration, and the cognitive costs associated with validating and contextualizing their outputs.

When evaluating potential negative effects of intensive AI use, respondents express moderate but notable concern. Nearly 30\% worry that over-reliance may erode practitioners' understanding of Scrum processes, while concerns about blurred accountability (23\%), reduced human connection (26\%), and diminished engagement (17\%) reflect broader anxieties about the human dimension of Agile work. Neutral responses were relatively common, suggesting that these risks are perceived as contingent: they depend on how the tools are used, by whom, and under which constraints.


\begin{tcolorbox}[colback=gray!5, colframe=black!50, title=\faLightbulb\quad \textbf{Summarized Answer - Risks and Challenges (RQ4)},  rounded corners,  arc=3pt, boxsep=2pt,
  boxrule=0.4pt]
\scriptsize
Although LLM assistants are widely used in Scrum practice, their adoption is accompanied by persistent concerns about reliability and trust. Practitioners report frequent issues with near-correct but flawed outputs, inconsistent response quality, and privacy or confidentiality risks, which reinforce the need for human oversight. Contextual and organizational frictions, such as limited tool integration and challenges in adapting AI outputs to real project settings, further constrain effective use. Overall, the findings indicate that responsible adoption depends not only on individual judgment but also on organizational governance to mitigate risks while preserving accountability and team engagement.
\end{tcolorbox}

\section{Implications}
\label{sec5-implications}

The findings indicate that LLMs are already embedded in the daily routines of Brazilian Scrum practitioners, delivering clear gains in productivity and cognitive support. However, there are different concerns about reliability, contextual fit, and confidentiality remain. These perceptions point to practical, organizational, and research implications that must guide responsible adoption of LLM in Scrum management. 

For \textbf{\textit{practitioners}}, the results suggest that LLMs function effectively as cognitive aids for preparing and refining Scrum artifacts, exploring solution alternatives, and structuring reasoning during planning and analysis. However, the high incidence of ``almost correct'' outputs (81\%) and ``hallucinations'' (59\%) highlights that AI-generated content should remain subject to careful review. Teams must preserve human ownership over negotiation, prioritization, and decision-making, ensuring that reliance on assistants does not erode accountability or engagement. For \textbf{\textit{organizations}}, the prominence of confidentiality concerns (63\%) and legal or ethical doubts (56\%) indicates the need for explicit guidelines on the appropriate use of LLM assistants. Policies on data sharing, validation of generated content, and acceptable use would support responsible adoption. Training initiatives focused on prompt articulation and critical evaluation of outputs may further amplify benefits while reducing risks. Lastly, for the \textbf{\textit{researchers}}, the findings empirically reinforce the opportunities to investigate how hybrid human–AI workflows reshape collaboration, learning, and knowledge sharing within Agile teams. The results also highlight the importance of studying mechanisms to mitigate reliability issues and to adapt assistant outputs to specific contexts.

\section{Threats to Validity}
\label{sec6-threats}

As with any survey-based study, this research is subject to potential threats to validity. We describe these threats below, structured according to the common standards, and outline the measures taken to mitigate them. 

\textbf{\textit{Internal validity}}. This study relied on self-reported data, which may capture perceptions rather than actual practices, an inherent limitation of surveys. To mitigate this, the instrument emphasized concrete, practice-oriented items (e.g., frequency of LLM use, supported artifacts, impacted roles) to reduce overgeneralized responses. Response bias, including social desirability, was addressed by ensuring anonymity, voluntary participation, and clarifying that responses would not be individually evaluated. Non-response bias may have occurred if professionals less familiar with LLMs chose not to participate; however, the survey was disseminated across diverse channels (professional networks, Agile communities, industry partners) to broaden reach and reduce systematic exclusion.

\textbf{\textit{External validity}}. The generalizability of the findings is limited by the use of non-probabilistic sampling strategies (convenience and snowball sampling). While these limitations represent a lack of representativeness, we mitigated the risk by targeting participants from diverse industries, roles, organizational sizes, and regions, thereby increasing heterogeneity. Another concern is temporal validity, as LLM technology evolves rapidly. Since the data were collected in October 2025, future adoption patterns may differ. We therefore clearly report the collection period and technological context to support proper comparison in future studies.

\textbf{\textit{Construct validity}}. These threats stem from the difficulty of accurately capturing perceptions of benefits, challenges, and expectations regarding LLM use in Scrum contexts. To strengthen construct validity, the survey design followed established methodological guidelines \cite{molleri2020empirically} and prior work on AI adoption. The instrument underwent expert review and pilot testing to refine wording, reduce ambiguity, and ensure alignment with the study objectives. Common method bias was mitigated by combining objective items (e.g., usage frequency, models adopted) with perceptual ones (e.g., benefits, risks) and by incorporating open-ended questions for qualitative triangulation. Because the study relied on a single survey instrument, mono-method bias remains a limitation; to address this, we recommend complementary approaches (such as interviews, focus groups, or case studies) in future work. Finally, potential language and interpretation bias was reduced through clear wording, explicit definitions of technical terms, and pilot validation with practitioners.

\textbf{\textit{Reliability}}. Reliability threats arise from potential inconsistencies in responses or in the interpretation of open-ended answers. To mitigate the first, all survey items were made mandatory in Google Forms, ensuring complete responses and reducing variability due to missing data. For the qualitative analysis, two researchers independently coded a subset of open-ended responses to strengthen inter-coder agreement, with disagreements resolved by discussion. Additionally, clear coding guidelines were established to minimize interpretive bias.

\section{Final Remarks}
\label{sec7-final}

This study aimed to characterize how Brazilian professionals are integrating LLMs into Scrum management activities. To achieve this objective, we conducted a survey with 70 practitioners, of whom 33 already reported active use of AI chat assistants in Scrum contexts. The findings show that Brazilian Scrum practitioners who use LLMs already demonstrate substantial familiarity with these tools: 85\% report intermediate or advanced proficiency and 52\% use them daily, typically in short sessions embedded in routine activities (RQ1). Adoption spans four major areas, but with different intensities: Exploring Scrum shows the strongest engagement (65–85\% planned or current use), artifact and event tasks form a substantial mid-tier (40–70\%), while broader management activities remain more cautious (20–50\%), especially in strategic or value-driven work. (RQ2). The main benefits include increased productivity (78\%), improved artifact quality (78\%), and significant reduction of manual effort (75\%), reinforcing the role of LLMs as cognitive and textual accelerators rather than autonomous agents (RQ3). At the same time, serious concerns persist: 81\% frequently encounter ``almost correct'' outputs, 59\% report hallucinations, and 63\% cite confidentiality risks, indicating that effective use requires careful validation and organizational safeguards (RQ4).

The study highlights important implications. Organizations should establish clear usage policies, foster AI literacy, and ensure that human judgment remains central to negotiation, prioritization, and decision-making processes. Practitioners, in turn, should treat AI outputs as drafts that require verification. As LLMs continue to evolve, balancing efficiency gains with reliability, privacy, and collaboration will be essential for their responsible adoption in Scrum management.

Future research could examine how these hybrid human–AI workflows evolve over time, especially regarding team learning, collaboration, and accountability. Longitudinal studies may reveal whether LLMs shift from preparatory roles to more integrated forms of participation in Agile activities. Additional work is needed to develop mechanisms that mitigate reliability issues, support contextual adaptation, and incorporate organizational knowledge into AI-assisted workflows.

\section*{Artifacts Availability}
\label{sec-artifact}

Supplementary materials available in: \url{https://github.com/Ilusinusmate/surveyxp26}.

\bibliographystyle{splncs04}
\bibliography{references}

\end{document}